\documentclass[aps,prl,twocolumn,psfig,showpacs,superscriptaddress]{revtex4}
\usepackage{amsfonts}
\usepackage{amsmath}
\usepackage{graphicx}
\usepackage{bm}
\usepackage{amssymb}
\usepackage{times}
\usepackage{dcolumn}
\usepackage{cases}
\usepackage{txfonts}

\begin{document}

\title{  Quantum phase transition, universality and scaling behaviors in the spin-1/2 Heisenberg model with
ferromagnetic and antiferromagnetic competing interactions on honeycomb lattice}
\author{Yi-Zhen Huang}
\affiliation{Theoretical Condensed Matter Physics and Computational
Materials Physics Laboratory, School of Physics, University of Chinese Academy of Sciences, P. O. Box 4588, Beijing
100049, China}
\author{Bin Xi}
\affiliation{Department of Physics and Beijing Laboratory of Opto-electronic Functional Materials $\&$ Micro-nano Devices, Renmin University of China, Beijing 100872, China}
\author{Xi Chen}
\affiliation{Theoretical Condensed Matter Physics and Computational
Materials Physics Laboratory, School of Physics, University of Chinese Academy of Sciences, P. O. Box 4588, Beijing
100049, China}
\author{Wei Li}
\affiliation{Department of Physics, Key Laboratory of Micro-Nano Measurement-Manipulation and Physics (Ministry of Education), International Research Institute of Multidisciplinary Science, Beihang University, Beijing 100191, China}
\author{Zheng-Chuan Wang}
\affiliation{Theoretical Condensed Matter Physics and Computational
Materials Physics Laboratory, School of Physics, University of Chinese Academy of Sciences, P. O. Box 4588, Beijing
100049, China}
\author{Gang Su}
\email[Corresponding author. ]{Email: gsu@ucas.ac.cn}
\affiliation{Theoretical Condensed Matter Physics and Computational
Materials Physics Laboratory, School of Physics, University of Chinese Academy of Sciences, P. O. Box 4588, Beijing
100049, China}

\begin{abstract}
The quantum phase transition, scaling behaviors, and thermodynamics in the spin-1/2 quantum Heisenberg model with antiferromagnetic coupling $J>0$ in armchair direction and ferromagnetic interaction $J'<0$ in zigzag direction on a honeycomb lattice are systematically studied using the continuous-time quantum Monte Carlo method. By calculating the Binder ratio $Q_{2}$ and spin stiffness $\rho$ in two directions for various coupling ratio $\alpha=J'/J$ under different lattice sizes, we found that a quantum phase transition from the dimerized phase to the stripe phase occurs at the quantum critical point $\alpha_c=-0.93$. Through the finite-size scaling analysis on $Q_{2}$, $\rho_{x}$ and $\rho_{y}$, we determined the critical exponent related to the correlation length $\nu$ to be 0.7212(8), implying that this transition falls into a classical Heisenberg O(3) universality. A zero magnetization plateau is observed in the dimerized phase, whose width decreases with increasing $\alpha$. A phase diagram in the coupling ratio $\alpha$-magnetic field $h$ plane is obtained, where four phases, including dimerized, stripe, canted stripe and polarized phases are identified. It is also unveiled that the temperature dependence of the specific heat $C(T)$ for different $\alpha$'s intersects precisely at one point, similar to that of liquid $^{3}$He under different pressures and several magnetic compounds under various magnetic fields. The scaling behaviors of $Q_{2}$, $\rho$ and $C(T)$ are carefully analyzed. The susceptibility is well compared with the experimental data to give the magnetic parameters of both compounds.

\end{abstract}


\maketitle

\section{\expandafter{\romannumeral1.} Introduction}

In the past few decades, there has been a great deal of interest in quantum antiferromagnets with alternating interactions, which could give rise to exotic quantum states and intriguing critical behaviors due to the competition between various interactions and external fields. In such systems, the quantum phase transition, universality and scaling behaviors are particularly interesting, and they have attracted much attention recently. In studies of quantum phase transition, the D-dimensional system is often mapped to a relevant (D+1)-dimensional classical system. By utilizing the (2+1)-dimensional nonlinear $\sigma$ model \cite{Chakravarty}, a quantum phase transition for the quantum Heisenberg model on a square lattice was mapped to a 3-dimensional thermal phase transition, which belongs to the so-called classical O(3) Heisenberg universality.

The universality hypothesis \cite{MEfisher,Griffiths,Kadanoff} states that all critical systems with the same dimensionality, symmetry of order parameter, and range of interactions are expected to share the same set of critical exponents and scaling functions. They do not depend on the microscopic details of the system \cite{Anastasios,campostrini}. It is shown that the quantum phase transition of the explicitly competing dimerized quantum Heisenberg model belongs to the O(3) universality \cite{troyer}. In addition, attempts have been made to find models that present exotic features at the critical region beyond the O(3) universality. Wenzel \textit{et al}. \cite{wenzel} argued that a novel transition occurs in the $J$-$J'$ model, with the critical exponent $\nu$ and $\eta$ smaller than the widely accepted standard values by quantum Monte Carlo (QMC) \cite{campostrini}. However, later recalculations on larger lattices and the critical exponent $\nu$ gave converse results \cite{yasuda,Jiang}.

A non-O(3) universality phase transition was discussed by Sandvik \cite{deconfine-sandvik}, in which the Heisenberg model with extra four-spin interactions experience a continuous ``deconfined'' quantum phase transition from the ordered Neel phase to another ordered phase -- a valence bond solid. The critical exponents $\nu$ and $\eta$ in this model differ dramatically from the O(3) universality, and the critical features described fulfill a previous theoretical prediction \cite{Senthil}.

On the other hand, as early as in 1998, a set of organic compounds -- 2-iodo, 2-bromo and 2-cyclopropy nitronyl nitroxide radicals (abbreviated as INN, BrNN, and $C_{3}H_{5}NN$) -- were found to form a spin-1/2 Heisenberg system on a honeycomb lattice with antiferromagnetic (AF) coupling $J$ along the armchair direction and ferromagnetic (F) coupling $J'$ along the zigzag direction, where the physical properties of INN and BrNN have been studied \cite{Hosokoshi}. Motivated by the experimental advances as well as previous theoretical considerations on honeycomb Heisenberg antiferromagnets \cite{Nakano,Shanks,Ugerber,Li}, it is interesting to ask if a quantum phase transition could occur at a critical coupling ratio $\alpha=J'/J$ in the spin-1/2 Heisenberg magnet with alternating AF and F interactions on honeycomb lattice, and what are the universality and scaling behaviors in this model. We shall answer these questions in this paper.

By using the continuous-time worldline QMC with worm update algorithm \cite{prokofef,Xi}, we showed that in this present system the spin stiffness $\rho$ scales as $\rho\sim L^{2-d-z}$\cite{fisher}, where $L$ is the lattice size, $d$ is the dimension of a system and $z$ is the dynamical critical exponent; the Binder ratio $Q_{2}$ of order parameter is a universality-related constant at the critical point. By means of these two quantities, the quantum critical point $\alpha_{c}$ and the critical exponent $\nu$ were calculated by $\rho_{x(y)}L_{x(y)}$ vs $\alpha$ and $Q_{2}$ vs $\alpha$ for different lattice sizes and finite-size scaling analysis\cite{Suzuki}, respectively, giving rise to $\alpha_{c}= -0.93$, and $\nu=0.7212(8)$, which indicates that this system with alternating interactions falls into a classical O(3) universality class. A phase diagram of the present system is presented in which the phase boundaries between four different phases are identified. We also observed that the specific heat curves intersect at the ``isosbestic point'' $T^{*}/J=0.741192$, which is similar to the cases in other systems \cite{Schlager,Krasovitsky,Janoschek,Huang}. The susceptibility is well compared with the experimental data to give the magnetic parameters of both compounds.

This paper is organized as follows. In Sec. \uppercase\expandafter{\romannumeral2}, we give a brief introduction to the model and the definitions of relevant physical quantities. In Sec. \uppercase\expandafter{\romannumeral3}, we mainly discuss the quantum critical behavior and the features in the presence of a magnetic field. The thermodynamic characteristics of the system are shown in Sec. \uppercase\expandafter{\romannumeral4}. A conclusion is presented in Sec. \uppercase\expandafter{\romannumeral5}.

\section{\expandafter{\romannumeral2.} Model and Definitions}
\begin{figure}
\includegraphics[width=0.49\textwidth]{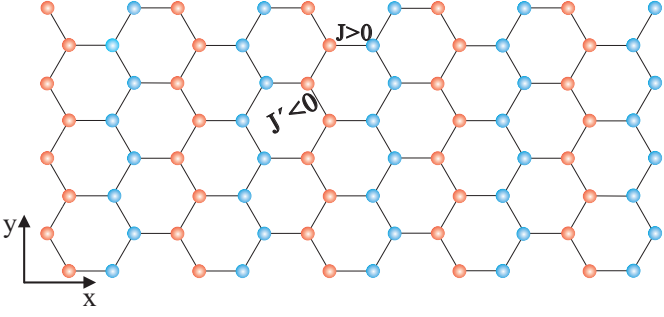}
\caption{ The spin 1/2 Heisenberg model on honeycomb lattice with anti-ferromagnetic (AF) and ferromagnetic (F) interactions along the armchair (X) and zigzag (Y) directions, respectively. This system is separated into sublattice A (red) and sublattice B (blue) and the lattice size are set to be $L_x=L_y=L$.   }
\label{fig-trans-field-diag}
\end{figure}
Let us consider the spin-1/2 Heisenberg model with F and AF competing couplings on honeycomb lattice, as shown in Fig. \ref{fig-trans-field-diag}, where the nearest neighbor spins couple via F and AF interactions along the zigzag and armchair directions, respectively. The Hamiltonian reads

\begin{eqnarray}
\begin{aligned}
H =J \sum_{\langle ij\rangle} \bold{S}_{i} \cdot \bold{S}_{j}+J' \sum_{\langle mn\rangle} \bold{S}_{m} \cdot \bold{S}_{n}-\emph{h} \sum_{l}{S_{l}^{z}},
\end{aligned}
\label{eq-hamiltonian}
\end{eqnarray}
where $\bold{S}_i$ is the spin-1/2 operator at the \textit{i}-th site, $J>0$ is the AF interaction along the armchair direction, $J'<0$ is the F interaction along the zigzag direction, $\emph{h}$ is an external magnetic field, $\langle ij\rangle$ means the summation over the nearest-neighbor sites along the x direction, and $\langle mn\rangle$ stands for the nearest neighbors along the y direction. We choose $J$ as the energy scale, and we presume the same lattice sizes along x and y directions, $L_{x}=L_{y}=L$.

We now utilize the worm QMC algorithm to explore the critical properties of this system. In the continuous-time path-integral representation, an imaginary time is introduced and the partition function $Z$ can be expressed as
\begin{eqnarray}
\begin{aligned}
Z&=Tr(e^{-\beta H})\\
&=Tr(e^{-\int_{0}^{\beta}d\tau H})\\
&=\sum_{n=0}^{\infty}(-1)^{n}\!Tr\{e^{-\beta H_{0}}\int_{0}^{\beta}\!d\tau_{n}...\int_{0}^{\tau_{2}}\!d\tau_{1}(H_{In}H_{I(n-1)}..H_{I1})\},
\end{aligned}
\label{partition-function}
\end{eqnarray}
where the Hamiltonian is separated into two parts $H_{0}$ and $H_{I}$ with $H_{Ii}=e^{\tau_{i}H_{0}}H_I e^{-\tau_{i}H_0}$, and
\begin{eqnarray}
\begin{aligned}
H_{0}=J\sum_{\langle ij\rangle}S_{i}^{z}S_{j}^{z}+J^{\prime}\sum_{\langle mn\rangle}S_{m}^{z}S_{n}^{z},
\end{aligned}
\label{diagonale-part}
\end{eqnarray}
\begin{eqnarray}
\begin{aligned}
H_{I}=\frac{1}{2}J\sum_{\langle ij\rangle}(S_{i}^{+}S_{j}^{-}+\emph{h}.\emph{c}.)+\frac{1}{2}J^{\prime}\sum_{\langle mn\rangle}(S_{m}^{+}S_{n}^{-}+\emph{h}.\emph{c}.).
\end{aligned}
\label{offdiagonal-part}
\end{eqnarray}
  By inserting the identity $\sum_{m}|\psi_{m}\rangle\langle\psi_{m}|=1$ between the operators in right-hand side of Eq. (\ref{partition-function}), where $\{|\psi_{m}\rangle\}$ is a complete set of basis and the single one of them is the direct product of eigenstates (spin up or spin down) of the local $S^{z}$, one may observe that the integrand in Eq. (\ref{partition-function}) can be viewed as the sampling weight. The Hamiltonian conserves the total $S^{z}$ of the system, so the sampling weight could be depicted as continuous lines of up (or down ) spins along the space-time representation, which is a worldline configuration \cite{Yasuda}. A worldline with two discontinuities (worm), which is a pair of upper and lower operators, is introduced into a $Z$ weight configuration. The two ends randomly walk in the space-time and destroy each other. When the two meet again and annihilate each other with a probability, a new $Z$ weight configuration is updated and sampled. More details about the updating scheme can be found in Ref. [\onlinecite{prokofef}]. Next, we will introduce several quantities that will be used to extract useful information for our purpose.

\subsection{\uppercase\expandafter{a.} Spin stiffness}

Spin stiffness $\rho$ is defined by the free energy $f$ and the boundary twist angle $\Phi$, $\rho\!=\! \partial^{2} f/\partial^{2} \Phi$ \cite{Wang,sandvik}, which is believed to scale as $\sim L^{2-d-z}$ near a quantum critical point. The dynamical exponent $z$ is presumed to be 1 at the beginning of calculations for this two-dimensional (2D) system. The quantities $\rho L$ in the x (armchair) and y (zigzag) directions are supposed to be size-irrelevant at the critical point, which can be calculated from the fluctuation of winding numbers in the stochastic series expansion (SSE) and worm algorithm
\begin{eqnarray}
\rho &=& \frac{3}{2 \beta}\langle W_{x}^{2}+W_{y}^{2}\rangle,
\label{eq-spinstiffness0}
\end{eqnarray}
where $\beta$ is the inverse temperature, $W_{x}$ and $W_{y}$ are the winding numbers in x and y directions, respectively, and $W_{\theta}$ in the $\theta$ direction is calculated by
\begin{eqnarray}
W_{\theta} &=& (N_{\theta}^{+}-N_{\theta}^{-})/L,
\label{eq-windingnu}
\end{eqnarray}
where $N_{\theta}^{+}$ ($N_{\theta}^{-}$) is actually the displacement of a down spin along the positive (negative) $\theta$ direction through the nearest neighbor jumping $S^{+}_{i+1}S^{-}_{i}$   ($S^{-}_{i+1}S^{+}_{i}$). Since the spin rotational symmetry are not broken in a finite-size system,  and because the symmetry breaking will not be restricted in the x-y plane in the thermodynamic limit,the pre-factor 3/2 is necessary in calculations in order to obtain the correct result. For the same reason, the total susceptibility and spin order parameter square \cite{Reger} should be the corresponding quantities in z (or x, y) direction multiplied by a factor 3. In boson systems, one can calculate the superfluid density $\rho_{s}$ by Eq. (\ref{eq-spinstiffness0}), which is also used to check the phase-transition point \cite{Yao} in a three-dimensional Bose-Hubbard model. In this present system, both $\rho_{x}L_{x}$ and $\rho_{y}L_{y}$ are measured at various $\alpha$ for different lattice sizes $L$.

\subsection{\uppercase\expandafter{b.} Binder ratio $Q_{2}$ }

The magnetic order parameter's moments near a critical point usually scales as \cite{beach,binder,kbinder,DPlandau}:
\begin{eqnarray}
\langle|m_{z}|^{k}\rangle_{L}=L^{-k\beta' /\nu}M_{k}(tL^{1/\nu}),
\label{moments}
\end{eqnarray}
where $L$ is the width of a lattice, $\beta'$ is the exponent for the order parameter (we here take the staggered magnetization $m_s \!\propto\! |1-T/T_{c}|^{-\beta'}$), and $\nu$ is the exponent for the correlation length. Binder ratio is defined by the ratio between two magnetization momenta
\begin{eqnarray}
Q_{k}=\frac{\langle m^{2k}\rangle_{L}}{\langle m^{2}\rangle^{k}_{L}},
\label{eq-ratios}
\end{eqnarray}
where $k$ is a positive integer, and the factor $L^{-2k\beta'/\nu}$ in the numerator and denominator cancel each other, leaving just the ratio of the scaling functions in the vicinity of the critical point. So, the Binder ratio is a dimension-free quantity for determining the critical point, and it is a universality-related constant \cite{beach}. We shall employ the second-order Binder ratio given by
\begin{eqnarray}
Q_{2}=\frac{\langle m_{sz}^{4}\rangle}{\langle m_{sz}^{2}\rangle^{2}},
\label{eq-binder}
\end{eqnarray}
where
\begin{eqnarray}
m_{sz}=\frac{1}{N}\sum_{1}^{N}\epsilon_i S_{i}^{z}
\label{staggered-magnetization}
\end{eqnarray}

\begin{eqnarray}
\langle m_{sz}^{2(4)}\rangle=\frac{1}{t M}\sum_{\eta=1}^{M}\sum_{\theta=1}^{t}(\frac{1}{N}\sum_{1}^{N}\epsilon_{i} S_{i(\eta\theta)}^{z})^{2(4)},
\label{staggered-magnetization-squared}
\end{eqnarray}
$N$ is the total number of lattice sites, and $\epsilon_{i}\!=\!+1$ $(-1)$ for sites on sublattice A (B), as indicated in different colors in Fig. \ref{fig-trans-field-diag}.
Eq. (\ref{staggered-magnetization}) gives the definition for the $z$ component of the staggered magnetization in the present case. For Monte Carlo algorithm, the order parameter's square is usually calculated by $\langle m_{s}^{2}\rangle=3\langle m_{sz}^2\rangle$.

In order to measure $\langle m_{sz}^{2}\rangle$ precisely, $M$ consecutive partition function configurations are sampled, in each of which $t$ time slices are picked out regularly. Every slice is actually a patten of Ising spin's arrangement, and a whole partition function configuration could be treated as some patten's continuous evolution along the imaginary time and back to itself in the end. The Worm update algorithm samples the most possible configurations, namely the pathes in Feynman theory, and treats the selected ones equally. For the calculation of $\langle m_{sz}^{2}\rangle$ (or $\langle m_{sz}^{4}\rangle$), all the time slices are treated with equal weights, and the arithmetic average over the staggered magnetization square (or quartic) is performed for all the sampled slices.

\subsection{\uppercase\expandafter{c.} Thermodynamic quantities }

 The average energy per site $\langle E\rangle$, magnetization $m$, susceptibility $\chi_{u}$ and the staggered magnetization $m_{\bot}^{s}$ in x-y plane of the system can be expressed by
\begin{eqnarray}
 \begin{aligned}
 \langle E\rangle=\frac{1}{\beta}(\langle\int_{0}^{\beta} U(\tau)d\tau\rangle+\langle N_{kinks}\rangle),
 \end{aligned}
 \label{eq_energy}
 \end{eqnarray}
 \begin{eqnarray}
 \begin{aligned}
 m=\frac{1}{L_{x}L_{y}}\sum_{i=1}^{N}\langle\frac{\int_{0}^{\beta}d\tau S_{i}^{z}(\tau)}{\beta}\rangle,
 \end{aligned}
 \label{eq_magnetization}
 \end{eqnarray}
 \begin{small}
 \begin{eqnarray}
\chi_{u}\!=\!\frac{1}{L_{x}L_{y}\beta}\!\sum_{ij}\!\int\! d\tau_{1}d\tau_{2}(\langle S_{i}^{z}(\tau_{1})S_{j}^{z}(\tau_{2})\rangle-\langle S_{i}^{z}(\tau_{1})\rangle\langle S_{j}^{z}(\tau_{2})\rangle),
\label{eq_uni_susceptibility}
\end{eqnarray}
\end{small}
\begin{eqnarray}
\langle (m_{\bot}^{s})^2\rangle=\frac{1}{N}\sum_{r=0}^{r=R_{max}}f_{r}(r)g^r(0,r),
\label{staggered-xy-plane}
\end{eqnarray}
where $U(\tau)=\sum_{ij}S_{i}^{z}(\tau)S_{j}^{z}(\tau)$ is the total nearest neighbor interactions in the z direction at imaginary time $\tau$, $N_{kinks}$ is the number of kinks in one sampled partition function configuration, $\beta$ is the inverse temperature, which acts as the length of imaginary time,
$g^r(t,r)=\langle\langle S^{-}{(t,r)};S^{+}{(0,0)}\rangle\rangle$ is the Green's function, a byproduct of the worm algorithm, and $f_{r}(r)= 1$ $(-1)$ if the two sites belong to the same sublattice (different sublattices).

In the absence of the external magnetic field, the susceptibility should be multiplied by a factor of 3, which can be calculated through the dynamic structure factor
\begin{eqnarray}
\chi_{u}=3S_{d}(0,0,0),
\label{eq_uni_susceptibility}
\end{eqnarray}
where the dynamic structure factor $S(q_{x},q_{y},\omega)$ is given by
\begin{eqnarray}
S_{d}\!=\!\frac{1}{L_{x} L_{y} \beta}\! \sum_{\langle ij\rangle}\!\int_{0}^{\beta}\!dtdt'\!e^{-i\vec{q}\cdot(\vec{r_{i}}-\vec{r_{j}})-i\omega (t-t')}\langle S_{i}^{z}(t)S_{j}^{z}(t')\rangle.
\label{eq-structure}
\end{eqnarray}
The specific heat can be obtained by \cite{Alessandro}
\begin{eqnarray}
C_{\nu}&=&\frac{1}{N}(\langle\beta^{2}E^{2}-N_{kinks}\rangle-\beta^{2}\langle E\rangle^{2})
\label{eq-specificheat}
\end{eqnarray}

The following simulations were carried on CPU E5620 with a frequency of 2.40 Ghz. For the cases in the ground state, the inverse temperature $\beta$ is set to be 100 for various lattice sizes, and the convergence for the case with $L_{x}=L_{y}=48$, $\beta=100$, $|\alpha|=0.932$ takes the most CPU time, about $ 10^{7}$ s.

\section{\expandafter{\romannumeral3.} Quantum phase transition and scaling behaviors}

\begin{figure}
\setlength{\belowdisplayskip}{3pt}
\includegraphics[width=0.5\textwidth,clip]{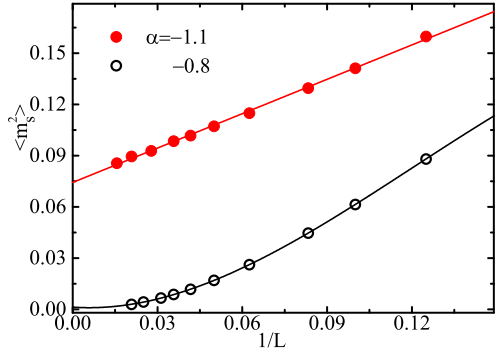}
\caption{(Color online) Inverse size extrapolation of order parameter $\langle m_{s}^{2}\rangle$ for $\alpha=-0.8$ and $\alpha=-1.1$ in the absence of a magnetic field. The intercept on $\langle m_{s}^{2}\rangle$ axis increasing from zero to about 0.07433 with increasing $|\alpha|$ indicates that a quantum phase transition from a magnetic disordered state to an ordered state might happen in the range of [-0.8,-1.1].}
\label{order-parameter}
\end{figure}

\subsection{\uppercase\expandafter{a.} Quantum critical point and scaling analysis}

\begin{figure}
\setlength{\belowdisplayskip}{3pt}
\includegraphics[width=0.5\textwidth,clip]{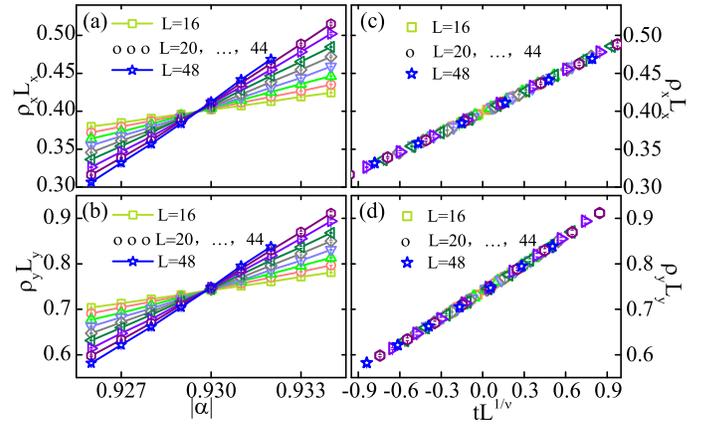}
\caption{(Color online) (a) $\rho_{x}L_{x}$ and (b) $\rho_{y}L_{y}$ as functions of $|\alpha|$ for lattice sizes L= 16, 20, 24, 28, 32, 36, 40, 44 and 48. For a given L, $\rho_{x}L_{x}$ and $\rho_{y}L_{y}$ increase linearly with increasing $\alpha$, and the two sets of curves have an intersection point at $|\alpha|\simeq 0.93$, indicating the existence of a QPT. Parts (c) and (d) are the corresponding data collapse by FSS analysis, where all curves drop onto a single straight line for each case. The errors for $\rho_{x}L_{x}$, $\rho_{y}L_{y}$ and $Q_{2}$ are at least three orders smaller than the corresponding quantities.}
\label{stiffness}
\end{figure}

\begin{figure}
\includegraphics[width=0.5\textwidth,clip]{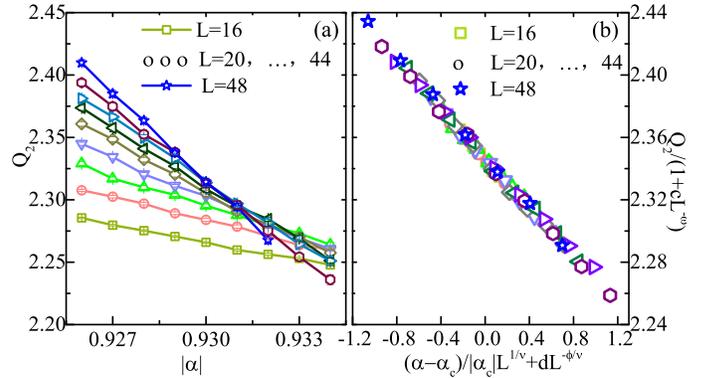}
\caption{(Color online) (a) The Binder ratio $Q_2$ as a function of $|\alpha|$ for different lattice sizes. The crossing points of curves are close to $|\alpha| \simeq 0.93$. (b) The scaling behavior of $Q_2$, where the corresponding scaling function contains a higher-order correction. }
\label{binder-ratio}
\end{figure}

To explore whether a quantum phase transition (QPT) exists in this system, we calculated the order parameter $\langle m_{s}^{2}\rangle$ in the absence of a magnetic field. We found that the inverse lattice size extrapolation of $\langle m_{s}^{2}\rangle$ increases from zero for $\alpha=-0.8$ to a finite value 0.07433 for $\alpha=-1.1$, as shown in Fig. \ref{order-parameter}. This shows that there might be a QPT from a magnetic disordered phase to an ordered phase in the range of [-0.8,-1.1]. To obtain the quantum critical point (QCP) accurately, we performed careful calculations on $\rho_{x}L_{x}$, $\rho_{y}L_{y}$ and Binder ratio $Q_{2}$ for different lattice sizes L= 16, 20, 24, 28, 32, 36, 40, 44 and 48 with $\alpha$ in the range of [-0.926, -0.934]. The results are presented in Fig. \ref{stiffness} and Fig. \ref{binder-ratio}, respectively. One may see that $\rho_{x}L_{x}$ and $\rho_{y}L_{y}$ increase linearly with increasing $|\alpha|$ for a given lattice size. Both  $\rho_{x}L_{x}$ and $\rho_{y}L_{y}$ curves for different lattice sizes have an intersection point at $|\alpha| \simeq 0.93$, indicating that it may be a QCP. Binder ratio $Q_{2}$ decreases with increasing $|\alpha|$ for a fixed lattice size, and  the curves for different lattice sizes do not intersect exactly at the same point (Fig. \ref{binder-ratio}), but these crossing points are very close to $|\alpha| \approx 0.93$, which implies that a higher-order correction should be included in the scaling functions according to the finite-size scaling hypothesis which will be discussed below.

\begin{table}
\renewcommand{\arraystretch}{1.5}
\addtolength{\tabcolsep}{+9pt}
\caption{ Critical coupling ratio $\alpha_{c}$ and the exponent $\nu$ of correlation length from the finite-size scaling analysis on spin stiffness and Binder ratio $Q_{2}$. The quantum phase transition is ascertained to happen at $\alpha_{c}\simeq -0.93$, which falls into an O(3) universality, as the exponent $\nu$ is close to 0.7112(5) of the standard classical Heisenberg O(3) universality.  }
\begin{tabular}[t]{c|ccc}
\hline \hline
 & $\rho_{x}L_{x}$ & $\rho_{y}L_{y}$  & $Q_{2}$ \\
\hline
$\alpha_{c}$ & -0.929526(4) & -0.929811(3)  & -0.93026(6)\\
$\nu$& 0.6748(8) & 0.7212(8) & 0.697(3) \\
$q_{0}$& 0.39908(2) & 0.74011(3) & 2.32(1)\\
\hline \hline
\end{tabular}
\end{table}

In light of the renormalization group theory \cite{Brezin,Barber,Zinn-Justin} and finite-size scaling hypothesis \cite{fisherfinite}, a quantity in a finite-size system, regardless of whether it is divergent or convergent at the critical point with a critical exponent $\kappa$, will obey a scaling function as long as the size L is large enough and the reduced phase-transition controlling parameter $t=(t-t_{c})/t_{c}$ is sufficiently small. In other words, the system is in a deep critical realm \cite{sandvik,beach}. In this present system, $t=(\alpha-\alpha_{c})/\alpha_{c}$. The scaling function has the form

\begin{eqnarray}
A(t,L)=L^{\kappa/\nu}g(tL^{1/\nu}),
\label{scaling}
\end{eqnarray}
where $\nu$ is the exponent of correlation length, $g(x)$ is a well-defined smooth function. At the critical point $t=0$, $g(x)$ is lattice independent, i.e., the curves of $A(t,L)/L^{\kappa/\nu}$ versus $tL^{1/\nu}$ for various size L would intersect at the critical point, as depicted in Figs. \ref{stiffness}. In fact, due to the lattice size accessible to the computation capacity and the nonlinearity of the scaling field, the curves of some quantities would not mutually intersect precisely at one point, e.g., the Binder ratio $Q_{2}$ in Fig. \ref{binder-ratio}. To utilize the data collapse method to calculate the critical exponents in this case, a higher-order correction should be included in the scaling function by
\begin{eqnarray}
A(t,L) = L^{\kappa/\nu}(1+cL^{-\omega})f(tL^{1/\nu}+dL^{-\phi/\nu}).
\label{scaling-correction}
\end{eqnarray}
For $\rho_{x}L_{x}$, $\rho_{y}L_{y}$ and $Q_{2}$, $\kappa$ is zero. Thus, we need to expand the scaling functions $g(x)$ and $f(x)$ as the second-order and forth-order polynomials of $tL^{1/\nu}$ and $tL^{1/\nu}+dL^{-\phi/\nu}$, respectively. The zero-order term in both cases is denoted by $q_{0}$.

To contrust a scaling analysis, the code of Melchert \cite{Melchert}, which can extract critical exponents without acquiring the detail of scaling functions, was first used to do a rough fitting and the cursory results were delivered to the next elaborate parameter fitting calculation. Thousands of nonlinear Levenberg-Marquardt optimization algorithm (LMOA) \cite{Gill}-based fittings of the bootstrap resamples of raw data should be performed; the details can be found in Ref. \cite{Wang}.

The refined $\alpha_{c}$, $\nu$ and $q_{0}$ with high accuracy are listed in Table \uppercase\expandafter{\romannumeral1}. One may see that the critical coupling ratio $\alpha_{c}$ extracted from three quantities is only different by $10^{-4}$, which gives rise to the QCP $\alpha_{c} \simeq -0.93$. Sparse data in the relative range of scaling variable $tL^{1/\nu}$ and $tL^{1/\nu}+dL^{-\phi/\nu}$ lead to discrepancies among the three $\nu$s, and determined by the lowest $\chi^{2}/d.o.f$[e.g., the weighted sum of squares residual per degree of freedom ($d.o.f$) of $\rho_{y}L_{y}$ for each single LMOA fitting],  the $\nu$ for the present system is found to be 0.7212(8), which is very close to the standard value 0.7112(5) of  classical Heisenberg O(3) universality. Our result is consistent with that of Hosokoshi \textit{et al}. \cite{Hosokoshi,Nakano}. Thus, the QPT falls in the O(3) universality class. In terms of the results shown in Fig. \ref{order-parameter}, the QPT occurs between the dimerized state with a gapful excitation and a striped state in that the spins along the zigzag rows are arranged parallel while the spins on two neighboring zigzag rows are aligned antiparallel, where the former is a disordered state, and the latter is an ordered state. In Ref. \cite{Mezzacapo}, we note that a different system, e.g., the spin-1/2 Heisenberg $J_{1}-J_{2}$ AF model on honeycomb lattice, was considered, and it was found that, when the coupling ratio $J_{2}/J_{1}$ increases from 0 to 1, the system goes to the following phases: for $J_{2}/J_{1}\leq 0.2$ it is in the N\'{e}el phase; for $J_{2}/J_{1}\geq 0.4$ it is in the collinear phase; and for the intermediate region it is in the disordered phase. This is quite different from the phases in the present system.

 \begin{figure}
 \includegraphics[width=1.0\linewidth,clip]{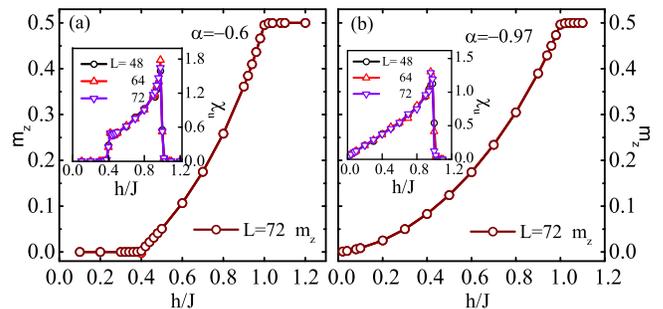}
 \caption{(Color online) $m_{z}$ and $\chi_{u}$ vs magnetic field $\emph{h}$ for different lattice size L=48, 64 and 72, where the susceptibility $\chi_{u}$ is plotted in the inset. (a) $\alpha=-0.6$; (b) $\alpha=-0.97$. With increasing $h$, for (a), the system goes through the dimerized phase (the region of zero magnetization plateau), the canted stripe phase, and the polarized phase successively, and for (b) it enters the canted stripe phase from the stripe phase directly. The discontinuities in $\chi_{u}$ indicate that all the QPTs are of second-order. The errors for susceptibility and magnetization are at least two and three orders smaller than the corresponding quantities, respectively.}
 \label{comparision}
 \end{figure}

\subsection{\uppercase\expandafter{b.} Phase diagram in a magnetic field}

Now let us consider the case in the presence of a magnetic field $h$. We calculated the magnetization per site and uniform susceptibility at $\alpha=-0.6$ and $\alpha=-0.97$ for lattice size L=48, 64 and 72. During the calculation, we found that the finite-size effect has no influence on the calculated results when the spin rotational symmetry is broken, and the discrepancies for the three lattice sizes are invisible, as indicated in Fig. \ref{comparision}.

\begin{figure}
 \includegraphics[width=1.0\linewidth,clip]{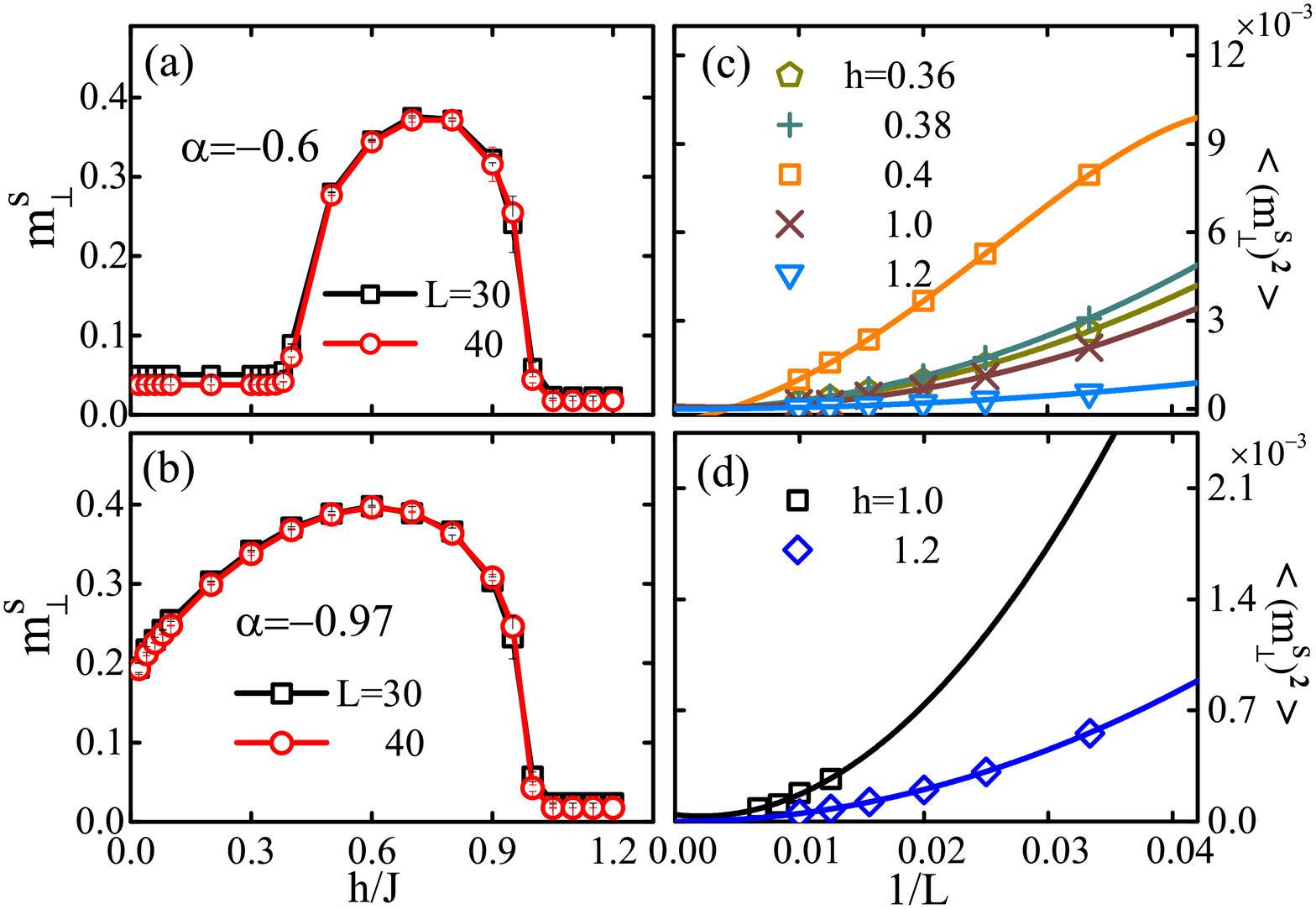}
 \caption{(Color online) Staggered magnetization in the x-y plane for (a) $\alpha=-0.6$ and (b) $\alpha=-0.97$. Parts (c) and (d) give the corresponding size extrapolations of $\langle (m_{\bot}^{s})^2\rangle$ for the dimerized and polarized phases with $\alpha=-0.6$ and $\alpha=-0.97$,  respectively, showing that in the thermodynamic limit there is no staggered magnetization in the x-y plane. Due to the tiny value in (c) and (d), the error bar for data is only shown in (a) and (b)}
 \label{size-extrapolation}
 \end{figure}

When $\alpha =-0.6$, there is a zero magnetization plateau in the magnetic curve, as shown in Fig. \ref{comparision}, which indicates that the system has a finite spin gap, suggesting that the system is in a dimerized state. We found that the width of the plateau decreases with increasing $\alpha$. The spin gap closes at the critical field $h{c1}=0.4$. When the magnetic field $h>h_{c1}$, the system enters into a new phase where the staggered magnetization $m_{\bot}^s$ in the x-y plane is nonvanishing (see Fig. \ref{size-extrapolation} (a)), giving rise to the canted stripe phase. In this phase, the uniform susceptibility $\chi_u$ increases with increasing field $h$ [the inset of Fig. \ref{comparision} (a)]. When the field exceeds the upper critical field, the system is polarized.

When $\alpha =-0.97$, there is no a zero magnetization plateau, implying that the low-lying spin excitation is gapless, and the magnetization increases with increasing the field and the staggered magnetization $m_{\bot}^s$ in x-y plane takes finite values below the saturation field [see Fig. \ref{size-extrapolation} (b)], indicating that the system is in a canted stripe phase. Again, in this phase, the uniform susceptibility increases from a finite value with increasing the field [the inset of Fig. \ref{comparision} (b)]. In both the dimerized phase and the polarized phase (PP), the size extrapolation shows that the squared staggered magnetization $\langle (m_{\bot}^{s})^2\rangle$ in the x-y plane vanishes in the thermodynamic limit, as manifested in Figs. \ref{size-extrapolation} (c) and (d) for $\alpha =-0.6$ and $\alpha =-0.97$, respectively.

When $h=0$, the system with $\alpha =-0.6$ goes into the dimerized phase due to the magnetization $m_z$, the staggered magnetization $m_{\bot}^s$ in the x-y plane, and the uniform susceptibility $\chi_u$ vanishing, as shown in Figs. \ref{comparision} (a) and \ref{size-extrapolation} (a); for $\alpha =-0.97$, the system is in the stripe phase in the absence of a magnetic field, as in this case, the magnetization $m_z$ and the uniform susceptibility $\chi_u$ are vanishing [Fig. \ref{comparision} (b)], but the staggered magnetization $m_{\bot}^s$ in the x-y plane is nonvanishing [Fig. \ref{size-extrapolation} (b)].

\begin{figure}
\includegraphics[width=1.0\linewidth,clip]{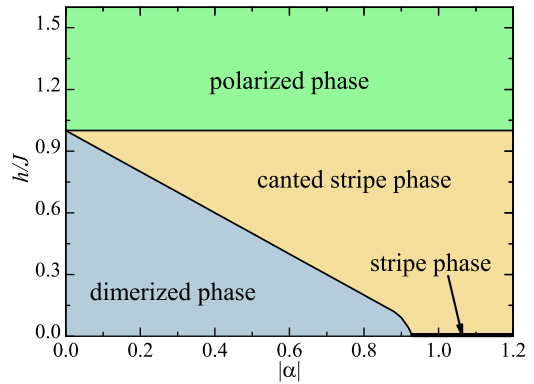}
\caption{(Color online) Phase diagram of the spin-1/2 Heisenberg model with F and AF alternating interactions on a honeycomb lattice in the coupling ratio $\alpha$-magnetic field $h$ plane. There are four phases: the dimerized, stripe, canted stripe and polarization phases. For $h=0$, there is a quantum phase transition (QPT) between the dimerized and stripe phase at the quantum critical point $\alpha_c=-0.93$. The QPT falls into the O(3) universality. All phase transitions in this system are of second-order. }
\label{phase-diagram}
\end{figure}

By summarizing the results extracted from the above calculations, the phase diagram of the present system in the $\alpha$-$h$ plane can be depicted, as shown in Fig. \ref{phase-diagram}. In the absence of the magnetic field, a small $\alpha$ gives a dimerized phase, because in this case the F interaction is weaker than the AF interaction, and the nearest-neighbor spins along the armchair direction form dimers, where the dimers are very weakly coupled through a single spin, enabling the system is in the dimerized state. A large $\alpha$ ($>\alpha_c$) gives a stripe phase, because in this situation the F interactions are strong enough to make spins along the zigzag direction form ferromagnetic chains which are coupled by anti-ferromagnetic interactions, leading the system to be in the stripe phase. When $\alpha=\alpha_c$, a QPT occurs between the dimerized phase and the stripe phase. By combining the scaling analysis on the spin stiffness and Binder's ratio, we show that the QPT is in the O(3) universality class.

In the presence of the magnetic field, when $\alpha < \alpha_c$, the dimerized phase remains under small fields, and when the field exceeds a certain value, the staggered magnetization in the x-y plane is nonvanishing besides a finite magnetization in z direction, leading to the system entering into the canted stripe phase; when $\alpha \geq \alpha_c$ the staggered magnetization in the x-y plane is always nonvanishing as long as the field is nonzero, implying that the system is in the canted stripe phase. When the field exceeds the saturation value, the system is fully polarized. By observing the values of $m_z$, $m_{\bot}^s$ and $\chi_u$, we determined the phase boundaries between different phases. From the uniform susceptibility, one may find that the phase transitions between various phases are of second-order.

\section{\expandafter{\romannumeral4.} Specific heat and susceptibility}


\subsection{\uppercase\expandafter{a.} Isosbestic point and scaling behavior of specific heat}

\begin{figure}
\includegraphics[width=1.0\linewidth,clip]{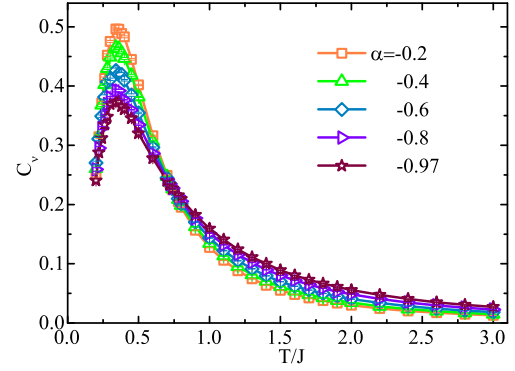}
\caption{(Color online) Temperature dependence of the specific heat for different $\alpha$ in the spin-1/2 Heisenberg model with F and AF alternating interactions on honeycomb lattice in the absence of magnetic field. A crossing point can be observed at $T^{*}/J\approx 0.74$. The error bars are invisible comparing with the symbol size.}
\label{specific-heat-different-alpha}
\end{figure}

The temperature dependence of the specific heat $C(T)$ of the spin-1/2 Heisenberg model with F and AF alternating interactions for different coupling ratio $\alpha$ in the absence of the magnetic field is studied, and the results are presented in Fig. \ref{specific-heat-different-alpha}. It can be seen that with increasing temperature, $C(T)$ increases first to a peak and then decreases to vanishing regardless of small or large $\alpha$, which is very similar to the bahavior of a spin-1/2 Heisenberg antiferromagnet. However, at low temperature, $C(T)$ has different behaviors for $\alpha < \alpha_c$ and $\alpha > \alpha_c$, where, when $T \rightarrow 0$, the former (in the dimerized phase) has an exponential decaying to zero, and the latter (in the stripe phase) has a power-law decaying to zero.

It is amazing to note that all curves of $C(T)$ for various $\alpha$ intersect at temperature $T^{*}/J\approx 0.74$. This phenomenon of the crossing point in specific heat in this present spin-1/2 Heisenberg model with F and AF alternating interactions on honeycomb lattice resembles that of liquid $^{3}$He, where the curves of specific heat $C_{\nu}(T,P)$ of liquid $^{3}$He under different pressures intersect precisely at $T_{+}\simeq 0.16K$. Such a crossing point $T^{*}$ in $C_{\nu}(T,\lambda)$, with $\lambda$ an intensive variable, has been reported in several magnetic and fermionic systems both theoretically and experimentally. For instance, the crossing point of $C_{\nu}(T,U)$ for different on-site repulsive interaction $U$ was theoretically argued to have a universal value for various Hubbard models by Vollhardt \textit{et al.} \cite{Chandra}; likewise, $C_{\nu}$ has a crossing point in several magnetic compounds such as $CeCu_{5.5}Au_{0.5}$ \cite{Schlager}, $RuSr_{2}Gd_{1.5}Ce_{0.5}Cu_{2}O_{10-\delta}$ \cite{Krasovitsky}, $MnSi$ \cite{Janoschek} and $CeAuSn$ \cite{Huang} under various external magnetic fields, respectively. There is a general argument for the crossing point given by Vollhardt \cite{Vollhardt}: since $\int_{0}^{\infty}dT\frac{C(T,|\alpha|)}{T}$ is the high-temperature limit of entropy S, it is independent of $|\alpha|$, as described in Eq.(3) in \cite{Vollhardt}, $\eta_{X}=k_{B}^{-1}\lim_{T\rightarrow\infty}\frac{\partial S(T,X)}{\partial lnX}=\frac{X}{K_{B}}\int_{0}^{\infty}\frac{dT^{\prime}}{T^{\prime}}\frac{\partial C(T^{\prime},X)}{\partial X}$. Consequently, there has to be one T-regime where the function takes positive values and one regime with negative values, leading to a unique T value at which the $\frac{\partial C_{v}(T,\lambda)}{\partial\lambda}|_{T^{*}}$ is zero. $T^{*}$ is the crossing point of the curves of specific heat, coined as ``isosbestic point". For the present spin system, $\lambda=\alpha$, $C_{\nu}(T,\alpha)$ intersects for various $\alpha$ at $T^{*}/J$ with $C_{\nu}|_{T^{*}} \approx 0.22$, as shown in Fig. \ref{specific-heat-different-alpha}.

\begin{figure}
\includegraphics[width=1.0\linewidth,clip]{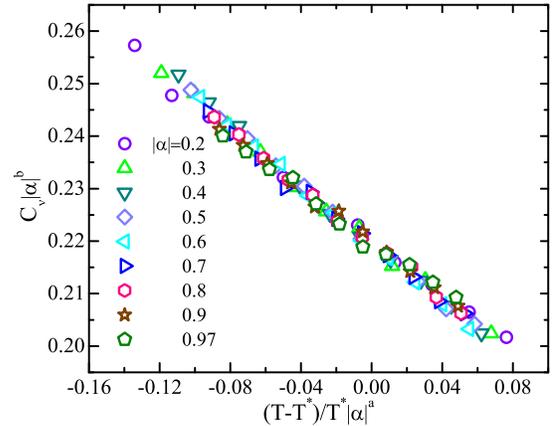}
\caption{(Color online) A scaling analysis for the specific heat of the spin-1/2 Heisenberg model with F and AF alternating interactions on honeycomb lattice in the vicinity of the ``isosbestic point '' $T^*=0.763(9)$. For various $\alpha$, all curves drop onto a straight line.}
\label{specific-heat-scaling}
\end{figure}

As the specific heat in this system has a crossing point at $T^*$ for various coupling ratio $\alpha$, there must be a scaling behavior in the vicinity of $T^*$. We propose the scaling function in the following
\begin{eqnarray}
C_{\nu}|\alpha|^{b}=q_{0}-b_{1}|\alpha|^{a}(1-\frac{T}{T^{*}}),
\label{scaling-spec-heat}
\end{eqnarray}
where $a$, $b$, $q_{0}$ and $b_1$ are constants. For $|\alpha|$ = 0.2, 0.3, 0.4, 0.5, 0.6, 0.7, 0.8, 0.9, 0.97 with L=64 in the temperature range of $[0.7,0.8]$, we plot $C_{\nu}|\alpha|^{b}$ versus $|\alpha|^{a}(\frac{T}{T^{*}}-1)$, and observe that all curves drop onto a straight line, with $T^{*}/J=0.763(9)$, $b=-0.019(5)$, $a=-0.29(7)$, $b_1=-0.267(3)$, and $q_{0}=0.2197(2)$, as shown in Fig. \ref{specific-heat-scaling}.

\begin{figure}
\includegraphics[width=1.0\linewidth,clip]{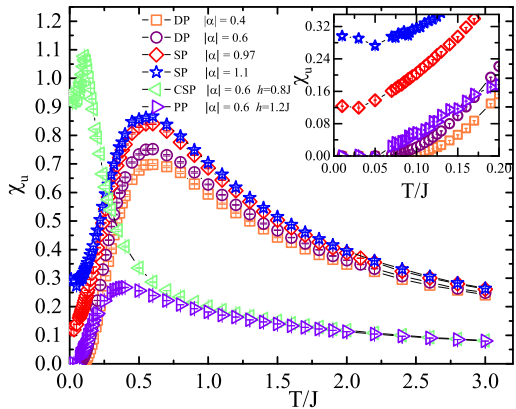}
\caption{(Color online) The temperature dependence of the uniform susceptibility $\chi_u$ in the spin-1/2 Heisenberg model with F and AF alternating interactions on honeycomb lattice for different $\alpha$. DP denotes dimerized phase, SP denotes stripe phase, CSP denotes canted stripe phase and PP denotes polarized phase. The lattice size is set to be L=64. The error bars could not be shown in the figure.}
\label{susceptibility-each-phase}
\end{figure}
\subsection{\uppercase\expandafter{b.} Susceptibility and comparison to experiments}

\begin{figure}
\includegraphics[width=0.5\textwidth,clip]{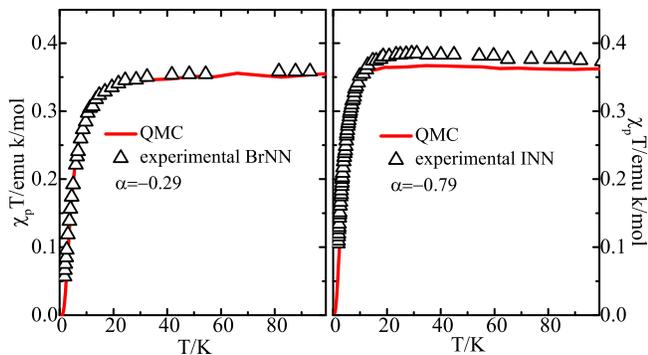}
\caption{(Color online) The paramagnetic susceptibilities for the organic radicals BrNN and INN are fitted to the model given by Eq. (\ref{eq-hamiltonian}) with the lattice size L=64. The fitting results give the coupling ratio $\alpha=-0.29$ and $-0.79$, and the AF coupling is $J=8.25 $ K and $6.9$ K for both materials, respectively, being well consistent with the experimental measurements \cite{Hosokoshi}. }
\label{susceptibility-th-ex}
\end{figure}

The temperature dependence of the uniform susceptibility $\chi_{u}(T)$ in the spin-1/2 Heisenberg model with F and AF alternating interactions on honeycomb lattice is explored for various coupling ratios $\alpha$, where the results are given in Fig. \ref{susceptibility-each-phase}. We found that in the absence of a magnetic field, at high temperature the susceptibility $\chi_{u}$ in the stripe phase (SP) exhibits a round peak higher than those in the dimerized phase (DP) and PP; at low temperature, $\chi_{u}$ in DP and PP decreases to zero very rapidly as $T$ decreases, while the one in SP keeps a finite value when $T\rightarrow0$. It is consistent with the fact that in the DP the low-lying excitation has a finite spin gap, leading to an exponential decaying of $\chi_{u}(T)$ at $T\rightarrow0$, while in the SP the spin excitation is gapless, resulting in a power-law decaying behavior of $\chi_{u}(T)$ at $T\rightarrow 0$, and in the PP with $\alpha=-0.6$, $\chi_{u}(T)$ is remarkably suppressed by the magnetic field $h/J=1.2$. In the canted stripe phase (CSP), $\chi_{u}(T)$ exhibits a quite different behavior from those three phases, in which a sharper peak of $\chi_{u}$ appears at lower temperature, and in the limit of $T\rightarrow0$ the system has a larger magnetic susceptibility in comparison with the SP case for $|\alpha|=1.1$. This is understandable, because in the CSP at $h/J=0.8$, the fact that the spins are partially polarized makes it easier to response to the external field with higher $\chi_{u}$, and the competition between the quantum fluctuations in the x-y plane and the thermal fluctuations results in such a peak in Fig. \ref{susceptibility-each-phase}.

The calculated behavior of susceptibility can be compared with the experimental data. Nearly two decades ago, two organic radicals BrNN and INN \cite{Hosokoshi} were reported to form a honeycomb lattice with F and AF competing interactions between nearest neighbor spins, which can be described by the model defined in Eq. (1). In comparison to the experimental observation, we fitted the experimental data of the uniform magnetic susceptibility $\chi_{u}(T)$ with the calculated results for the lattice size L=64, as shown in Fig. \ref{susceptibility-th-ex}. The fitting results give the coupling ratio $\alpha=-0.29$ for BrNN and $\alpha=-0.79$ for INN, with the AF coupling $J=8.25$ K and $6.9$ K, respectively. One may see that our results are well consistent with the experimental measurements \cite{Hosokoshi}.

\section{\expandafter{\romannumeral5.} Conclusion}

We have applied the continuous imaginary time quantum Monte Carlo method with worm update algorithm to study the spin-1/2 Heisenberg model with antiferromagnetic coupling in the armchair direction and ferromagnetic coupling in the zigzag direction on a honeycomb lattice, and we found that a quantum phase transition from the dimerized phase to the stripe phase occurs at the quantum critical point $\alpha_c=-0.93$. Through the finite-size scaling analysis on $Q_{2}$, $\rho_{x}$ and $\rho_{y}$, the critical exponent $\nu$ related to the correlation length is determined to be 0.7212(8), implying such a QPT falls into a classical Heisenberg O(3) universality. The dimerized phase exhibits a zero magnetization plateau, whose width decreases with increasing $\alpha$. A phase diagram in the coupling ratio-magnetic field is presented in which four phases, including the dimerized phase, the stripe phase, the canted stripe phase and the polarized phase are identified. The temperature dependence of the specific heat $C(T)$ shows an exotic feature, namely that the curves for different $\alpha$ intersect precisely at one point, similar to that of liquid $^{3}$He under different pressures and some magnetic compounds under different magnetic fields. The scaling behaviors of $Q_{2}$, $\rho_{x}$ and $C(T)$ are carefully addressed, and the scaling functions are given. The temperature dependence of susceptibility is explored systematically for the four phases, and it is also well compared with the experimental measurements. The magnetic parameters of the two compounds BrNN and INN are estimated, which are in good agreement with the experimental observation\cite{Hosokoshi,Nakano}. Our present study enables us to gain a deeper insight into the overall critical and noncritical nature of the spin-1/2 Heisenberg model with F and AF alternating interactions on honeycomb lattice, and it may help to guide future experimental studies.

\section{acknowledgement}

The authors appreciate discussions with Shi-Ju Ran, Xin Yan and Cheng Peng, and the help from Guangzhao Qin. This work was supported in part by the MOST of China (Grant  No. 2012CB932900 and No. 2013CB933401), the NSFC (Grant No. 14474279 and No. 11574200), the Beijing Key Discipline Foundation of Condensed Matter Physics (Grant No.11504014 ), and the Strategic Priority Research
Program of the Chinese Academy of Sciences (Grant No. XDB07010100).

\end{document}